\documentclass[prl,floatfix,twocolumn,amsmath]{revtex4}
\usepackage{graphicx}
\usepackage{color}

\begin{document}

\def\ket#1{|#1\rangle}
\def\bra#1{\langle#1|}
\def\av#1{\langle#1\rangle}
\def\myarrow{\mathop{\longrightarrow}}
\def\ua{\uparrow}
\def\da{\downarrow}
\setlength\abovedisplayskip{9pt}
\setlength\belowdisplayskip{9pt}
\setlength\belowcaptionskip{-8pt}


\title{All-Optical Production of quantum degeneracy and molecular BEC of $^6$Li }

\author{Shujing Deng, Pengpeng Diao, Qianli Yu,
and Haibin Wu$^{*}$ }
\affiliation{State Key Laboratory of Precision Spectroscopy, Department of Physics, East China Normal University, Shanghai 200062, China}
\email{hbwu@phy.ecnu.edu.cn}

\date{\today}

\begin{abstract}
We achieve a highly degenerate and strongly interacting Fermi gas  in a mixture of the two lowest hyperfine states of $^6$Li by direct evaporative cooling in a high power crossed optical dipole trap. The trap is loaded from a large atom number magneto-optical trap (MOT) which is realized by a laser system of 2.5-watts intracavity-frequency-doubled light output at 671 nm. With this system, we also demonstrate the production of a molecular Bose-Einstein condensates (mBEC), and observe the anisotropic expansion of Fermi gases in the so-called BEC-BCS crossover regime.
\end{abstract}
\maketitle


Ultracold Fermi gases have been a very active subject of many experimental and theoretical studies in past decades~\cite{Bloch,Giorgini}, which greatly impact many fields of physics from high temperature superconductors to neutron stars and quark-gluon plasmas~\cite{Johnnuclear}.  The controllability of the external trapping potentials and the interaction strength by Feshbach resonances (FR) has made them as ideal systems to study the exotic many-body phenomena. Such resonances characterize the two-body collisional interaction and permit one to change the value and even the sign of a scattering length by simply tuning an external
magnetic field and/or optical fields, therefore reaching the regimes of Bardeen-Cooper-Schrieffer (BCS) superfluid and Bose-Einstein condensation (BEC). The connection between two regimes is the strongly interacting BEC-BCS crossover where the scattering length divergences and the Fermi gases exhibit unitary properties~\cite{Zwerger}. By manipulating the external trap potential, reduced dimensional and interacting Fermi systems  can  be produced; one can therefore investigate the novel phase diagrams and simulate the condensed matter physics.

Since the first achievement of strongly interacting Fermi gases~\cite{Oharadenerancy}, the unprecedented progress on studying the degenerate Fermi gases (DFGs) has been achieved. It becomes very important to study such novelty many-body physics with the simplified and reliable systems nowadays. In this work, we report on the achievement of  quantum degeneracy of Fermi gases in a mixture of the two lowest hyperfine states of $^6$Li with an improved system. A laser system of 2.5-watts intracavity-frequency-doubled light output at 671 nm is used to realize a standard magneto-optical trap (MOT). Near $ 10^{10}$ atoms could be loaded from a 30 cm length Zeeman slower. Combined with a high power (200 W) crossed optical dipole trap, such large atoms  greatly increase the initial phase space density and facilitate the atoms cooling and trapping. The highly degenerate Fermi gas is realized near FR. With this system, an mBEC is demonstrated by direct forced evaporative cooling at BEC side.  We also investigate the anisotropic expansion of such strongly interacting Fermi gases on unitary regime. The described system represents the excellent starting point for the study of novel universal dynamics with a large-atom number quantum degenerate Fermi gas.

\begin{figure}[htb]
\includegraphics[width=3.5 in ,height=2.7 in]{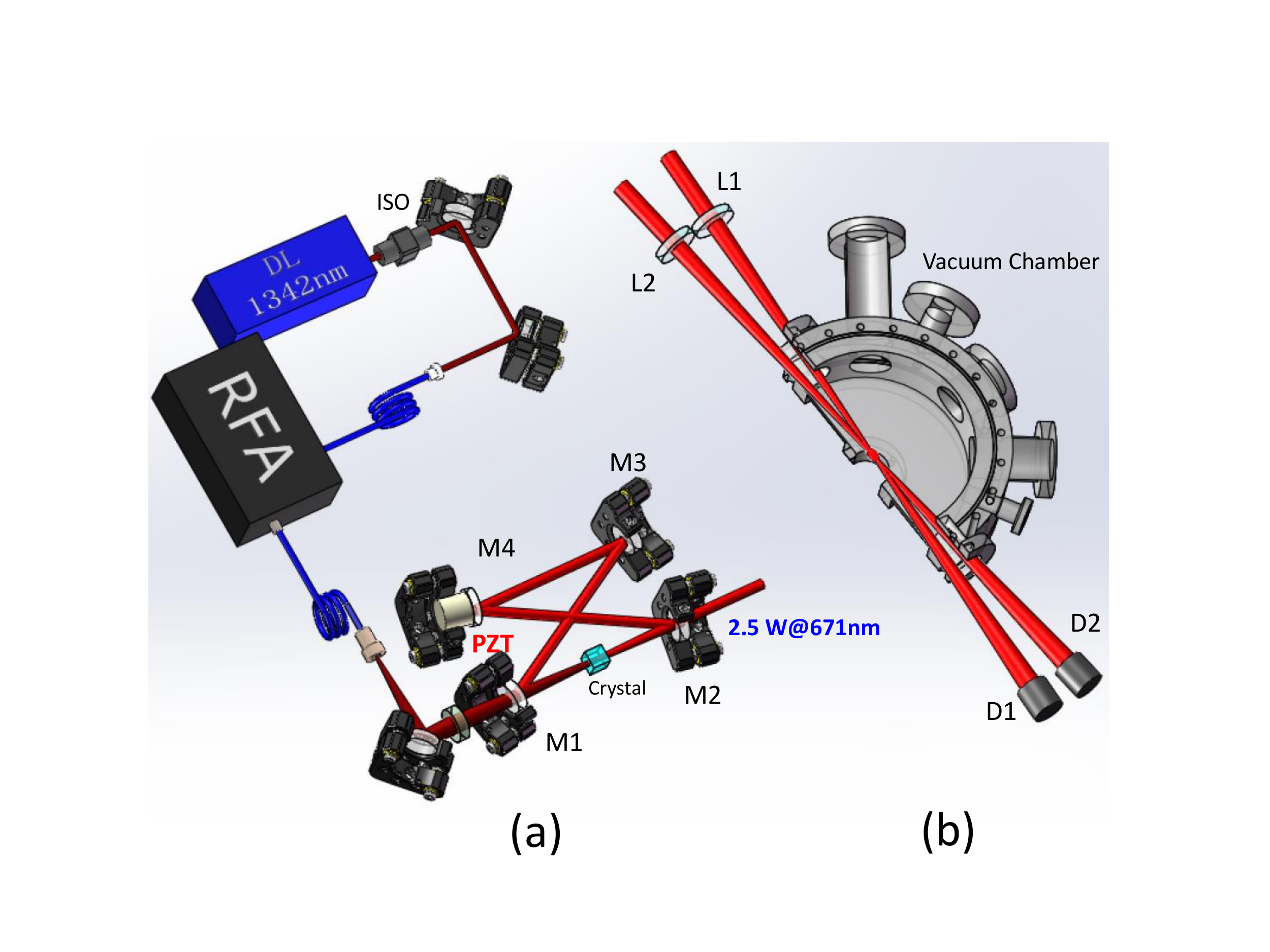}
\caption[example]
   { \label{fig:fig1}
The schematic of experimental setup.  (a) Laser setup for lithium MOT. RFA, Raman fiber amplifier; ISO, optical isolator; M1-M4, doubling-cavity mirrors; PZT, piezoelectric transducer. (b) The schematic of crossed dipole trap. L1-L2, achromatic lenses; D1-D2, dumpers.}
\end{figure}

The schematic of experimental setup for the cooling and trapping of fermionic $^6$Li atoms is shown in Fig. 1. A commercial extended cavity diode laser (Toptica DL Pro) with about 35mW output at 1342 nm is used as a seeding laser.  After an optical isolator and a single-mode optical fiber, about 25 mW is injected into a Raman fiber amplifier (RFA) and amplified to 5W, which is served  as pump source to an intracavity-frequency-doubler (LEOS). The conversion efficiency is about 50$\%$ and therefore 2.5 W laser light output at 671 nm is generated. The doubling cavity is locked by Hansch and Coullaud methods. Its frequency mode-hop tuning range is about 5GHz limited by the tunability of the used piezoelectric transducer (PZT). The frequency of such laser system is locked to red detuning 200MHz of the transition $2 S_{1/2},\,F=3/2\rightarrow 2 P_{3/2},\,F=5/2$ of $D_2$ line by the saturation absorption spectroscopy in an atomic lithium vapor cell. The port of laser output ($150\,mW$) is directly used as Zeeman slower beam. Others are frequency shifted by acousto-optic modulators (AOMs) to serve as the cooling beams and repumping beams. Typically $5.2\times 10^9$ atoms is loaded from a Zeeman slower for 10 sec. In order to get high initial phase space density, the two-stage cooling is adopted. After the loading stage the MOT are compressed, i.e., the cooling laser beams are tuned about 10 MHz below resonance and lowered in intensity to far smaller than saturation intensity and simultaneously the MOT gradient magnetic field is quickly shifted from 16 $G/cm$ to 50 $G/cm$. Finally cold atoms of near Doppler-limited temperature of 280 $\mu K$ and a density of $10^{12}$ cm$^{-3}$ are obtained. Following this compressed stage, the MOT gradient magnets are extinguished and repumping beams are switched off faster than the cooling beams. The $F=3/2$ hyperfine state is emptied to produce a $50-50$ mixture of atoms in the lower states $|1\rangle\equiv|F=1/2, M=-1/2\rangle$ and $|2\rangle\equiv|F=1/2, M= 1/2\rangle$.

Our optical dipole trap is realized  by a 200$\,$ W Ytterbium fiber laser at 1070$\,$nm (YLR-200-LP-AC), as shown in Fig1. (b). Two beams  are crossed at 19$^0$ at the centre of the MOT. The polarizations are orthogonal to avoid the interferences, which can cause to the substantial decrease of the trap lifetime. At the maximum laser powers of 80$\,$ W and 78$\,$ W of the beams at the focus, the radial and axial oscillation frequencies are measured by parametric resonances in the weakly interacting regime, which are $\omega_r=2\pi\times 26.5\, kHz$ and $\omega_z=2\pi \times 2.4\,kHz$, respectively. The full trap depth $U_0$ is about $k_B\times 4.8\,mK$ ($k_B$ is Boltzmann's constant).

\begin{figure}[htb]
\includegraphics[width=3.3 in]{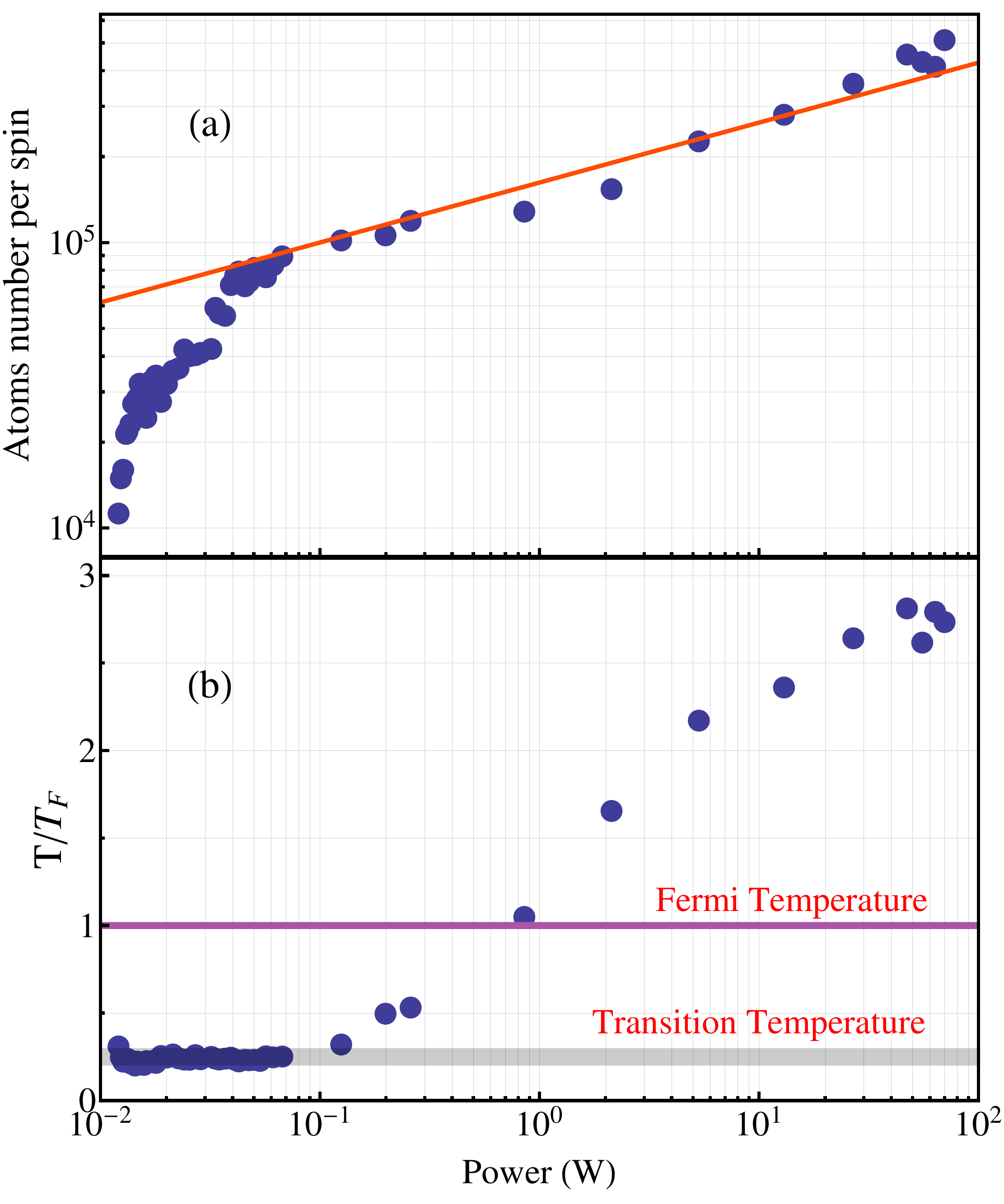}
\caption[example]
   { \label{fig:fig1}
Evaporative cooling to quantum degeneracy in the crossed optical dipole trap at Feshbach resonance $B=832$ G. (a) Atom number $N$ in state $|1\rangle$ versus the final dipole trap laser power. The blue dots are measured data and the solid red curve represents the scaling law predicted in Ref.~\cite{Scalinglaw}. (b) The dimensionless temperature $T/T_F$ as a function of the final dipole trap laser power. $T_F=\hbar \bar{\omega} (6 N)^{1/3}/k_B$ is the Fermi temperature of an ideal Fermi gas, where the mean trap frequency $\bar{\omega}=(\omega_r^2\omega_z)^{1/3}$. The purple line denotes the degeneracy temperature and the gray line represents the transition temperature for the superfluid. }
\end{figure}
The initial peak phase-space density is $3\times 10^{-3}$. The mean elastic collision rate is as high as $~1.7\times 10^5$. With such high trapping frequencies and tunably large scattering length, the evaporative cooling is easily preformed. The optical dipole trap is turned on 100ms before the MOT compressed stage. After the atoms are loaded into the dipole trap we hold the atoms 200 ms on the trap and then forced evaporative cooling is followed by lowering the trap laser power. We use a simple exponential ramp [$U(t)=U_0\exp(-t/\tau)$] as a lowering curve, where $U(t)$ is the final trap depth and the time constant $\tau=0.2-0.6$ s is selected and optimized by the different experimental requirements. The system allows us to precisely control the relative trap depth ($U(t)/U_0$) to levels below $10^{-4}$. After the forced evaporative cooling the ultracold atoms are held 0.5 s for the equilibrium  and then exponentially raised to the specific trap depth in 0.3 s for imaging. The image light comes from another diode laser (Toptica DL Pro). The frequency is locked into the cooling laser by a beatnote locking technique. An accurate oscillator allows us to tune the frequency detuning of the image light and cooling light from 250 MHz to 3 GHz, which greatly facilitates the imaging at high magnetic fields.

\textit{Evaporative cooling to quantum degeneracy}. Figure 2 is the results of evaporative cooling in the crossed optical dipole trap at Feshbach resonance $B=832$ G. Fig. 2(a) shows  atom number $N$ per spin (in state $|1\rangle$) as a function of the final dipole trap laser power. The measured atom number first follows a scaling law $N/N_0=(U(t)/U_0)^\alpha$~\cite{Scalinglaw}, with $\alpha=0.21$, as shown of red curve in Fig. 2(a). As decrease of the trap laser power the more energetic atoms are escaped from the trap. The atoms spilling effect~\cite{JochimScience} is observed when the trap depth is lowered close to the Fermi energy $E_F$. Since the trap depth decreased fast than  $E_F$, the atoms experiences a rapid loss at the very low trap depth. Fig. 2(b) shows the dimensionless temperature $T/T_F$ verse the final dipole trap laser power, where $T_F=\hbar \bar{\omega} (6 N)^{1/3}/k_B$ is the Fermi temperature of an ideal Fermi gas and the  mean trap frequency $\bar{\omega}=(\omega_r^2\omega_z)^{1/3}$. The method used to determine the temperature is the same as Ref.~\cite{Temperature1, Temperature2,Temperature3}.  The cloud size is first measured by an absorption image and then the virial theorem is used to obtain the energy of the strongly interacting Fermi gas~\cite{Virialtheorem}. The corresponding temperature is abstracted by its relation with the energy.  Fermi degeneracy, where the energy of the ultracold atomic gas reached Fermi energy $E_F$ (purple line in Fig. 2(b)), takes place at $U(t)/U_0=8.75\times 10^{-3}$.  As the evaporative cooling is further performed, the temperature of the gas keeps decreasing and reaches the critical temperature of the superfliud with about $2\times 10^5$ atoms at two spins. In our case, as the trap laser power is further decreased we don't see the apparent decrease of the temperature and on the contrary we see a little increase of the temperature of the gas. The mechanism of such the heating is not so clear currently and one possible reason is due to the losses of Fermions at the very shallow dipole trap depth~\cite{Carrheating}.

\textit{Molecular BEC}.  When the magnetic field is tuned to BEC side (positive scattering length) of FR and the temperature is below the molecular binding energy $E_B=\hbar^2/ma^2$, where $m$ is the atom mass and $a$ is s-wave scattering length, respectively, molecules are formed and remain trapped. Such molecules are composite bosons and can be condensed at sufficient low temperature, as reported previously in Refs.~\cite{JochimScience, BEC2,BEC3,BEC4,BEC5}. Fig. 3 shows the experimental observation of an mBEC of $^6$Li$_2$ in the crossed optical dipole trap. The gas is directly evaporative cooling at $B=810\,G$ ($a=17,126\,a_0$, $a_0$ is bohr radius). When the temperature is cooled near $E_B=k_B\times 100\,nK$, FR molecules are formed. At the end of the evaporation cooling the gas is strongly interacting. The scattering between the molecules is a four-body process and its scattering length is predicted to be on the same order as the atom-atom scattering length ($a_{mol}=0.6 a$)~\cite{BECStablility}. The relaxation into deep bound states of such fermionic dimers in contrast to their bosonic counterparts is strongly suppressed by the Fermi quantum statistic. To image the molecules cloud and better manifest the condensed properties, we ramp the magnetic field from 810 $G$ into 690 $G$ in 100 $ms$ and release the cloud from the trap. After the time-of-flight (TOF), the resonant absorption image is taken.

Fig. 3 shows a series of absorption images and the corresponding density distributions of the FR molecules at the weak trap axial axis with TOF=700$\mu s$ at $B=690\, G$ for different final evaporative cooling powers. The typical bimodal distributions of mBEC are observed, which show the evidence for the temperatures of the molecules well below the condensed temperature $T_c$. The green dots are the measured data. The wings of the profiles (thermal molecules) are fit to a Gaussian function (the blue curves). The red blue curves are fit to the combined profiles of Gaussian and Thomas-Fermi distribution. The thermal and condensed fractions as well as the corresponding temperatures can be abstracted from the fits. In the stage of the magnetic field ramp, the binding energy $E_B=k_B\times 16\mu K$ at $690\,G$ is a few times larger than the trap depth and the gas is almost consisted of the pure molecules.  When the final evaporative cooling power $P=66\, mW$ in Fig. 3(a), the condensed fraction (CF) is bout $19\%$ corresponding to $T=0.93\,T_c$. In Fig. 3(b), when $P=20\, mW$, $CF=46\%$ and $T=0.81\,T_c$. For $P=13\, mW$, one can get $CF=68\%$ and $T=0.68\,T_c$ (Fig. 3(c)), which $T_c=0.94 \hbar\bar\omega/k_BN_{mol}^{1/3}=512nK$.

\begin{figure}[htb]
\includegraphics[width=3.3 in]{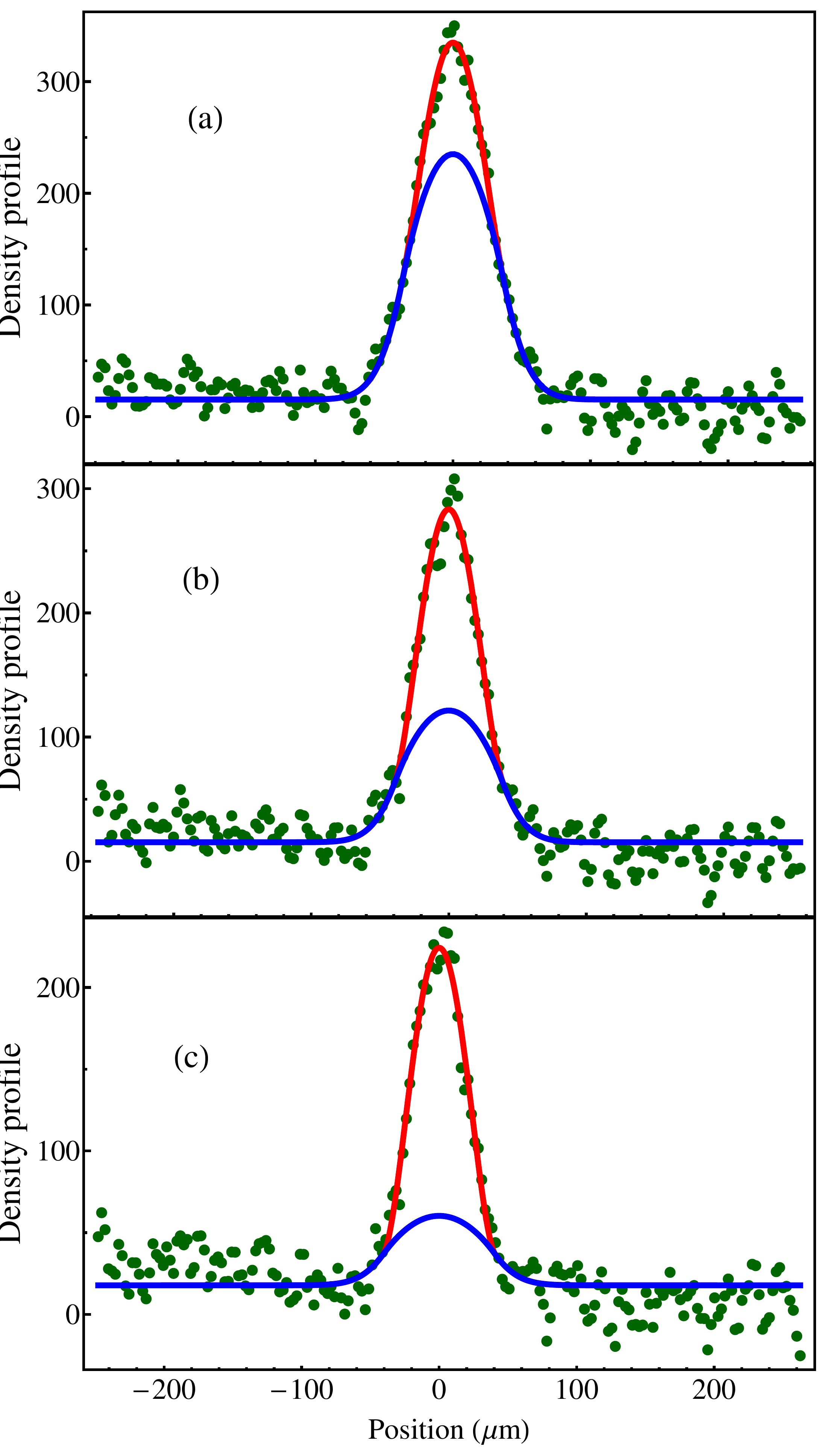}
\caption[example]
   { \label{fig:fig3}
The density profiles of $^6$Li$_2$ FR molecules along the weak trapping axial axis at the different final evaporative cooling powers. It clearly shows the transition to a molecular BEC. All images were taken with time of flight 700$\mu s$ at a magnetic field $B=690\,G$ after the forced evaporation at $B=810\,G$. The green dots are the measured data. The blue curves and the red blue curves are fit to Gaussian and the combined profiles of Gaussian and Thomas-Fermi, respectively. (a) $P=66\, mW$, the condensed fraction $CF=19\%$, $T=0.93\,T_c$. (b) $P=20\, mW$, $CF=46\%$ and $T=0.81\,T_c$. (c) $P=13\, mW$, $CF=68\%$ and $T=0.68\, T_c$. $T_c\equiv 0.94\hbar\bar\omega/k_BN_{mol}^{1/3}=512nK$ is the condensed temperature, which $N_{mol}$ is the molecules number. }
\end{figure}

\begin{figure}[htb]
\includegraphics[width=3.1 in]{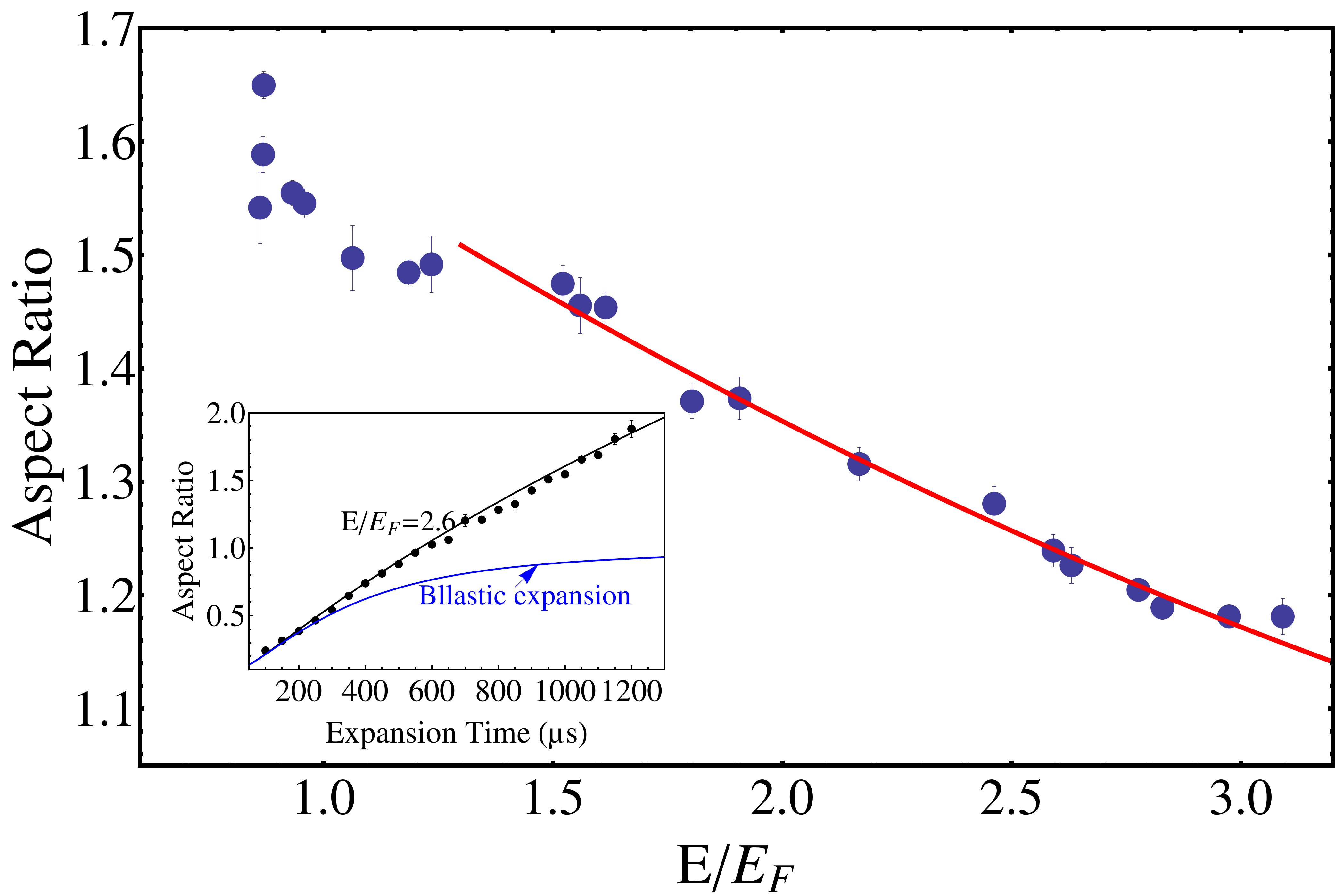}
\caption[example]
   { \label{fig:fig1}
Aspect ratio with expansion time 700$\mu s$ versus the dimensionless energy $E/E_F$ on unitary ($B=832.15 G$). The red curve is a quadratic fit at high energy regime. Inset: Aspect ratio as a function of expansion time with different TOF at $E/E_F=2.6$. The solid black curve is fit with the method used in Ref. \cite{CaoScience}. Error bars denote one standard deviation of the statistic. The blue curve is the theoretical ballistic expansion for an ideal Fermi gas. }
\end{figure}

After the switch off the trap, the strongly interacting Fermi gas experiences an elliptic flow  due to anisotropic pressure gradient at the weak and strong confined axes of the trap, which is first observed in Ref. ~\cite{Oharadenerancy}. Although such expansion confirming predictions of the hydrodynamic theory of superfluids at low temperature cannot be considered proof of superfluidity due to similar behavior existed in the collisional regime of a normal gas above the critical temperature, the study of the expansion dynamics is very important to understand on the strongly interacting Fermi gas. Recently, the elliptic flow of aspect ratio and ballistic expansion for the mean square cloud size have been used to measure the shear viscosity~\cite{CaoScience} and bulk viscosity~\cite{BulkPRL}. We also investigate such direct expansion of the strongly interacting Fermi gases on unitary regime. The aspect ratio (the ratio of cloud size of radial  axes and the elongated axial axes) at TOF=$700\,\mu s$  as a function of the dimensionless energy $E/E_F$ is shown in Fig. 4. Inset is the aspect ratio versus expansion time for different TOFs.  As the increase of energy, the expansion at radial direction becomes slow due to the increased viscosity and the aspect ratio smoothly evolves to the collisionless value at high energy. As the trap-averaged  shear viscosity shows the scaling of $(T/T_{F0})^{3/2}$ and $(T/T_{F0})\propto (E/E_F)^2$ at high temperature regime, where $T_{F0}$ is the local Fermi temperature at the trap center, the aspect ratio is related to $(E/E_F)^2$.  The fit with a quadratic function for high energy data is shown in Fig. 4, which agrees very well the experimental measurements. It is evidence that the aspect ratio at low energy greatly deviates from the fit, which is of particular interest and needs to be investigated further. The rigorous treatment could be derived from the dissipative hydrodynamics theory~\cite{CaoScience}.

In conclusion, we achieve quantum degeneracy of Fermi gases and the mBEC in a mixture of the two lowest hyperfine states of $^6$Li with an improved system. MOT with large atom numbers are realized by the laser system of 2.5-watts intracavity-frequency-doubled light output at 671 nm. Combined the high power crossed optical dipole trap and the large atoms greatly facilitate the atoms cooling. The strongly interacting degenerate Fermi gas is realized. An mBEC is demonstrated by direct evaporation at BEC side of FR in this system.  We also observe the anisotropic expansion of such strongly interacting Fermi gases on unitary regime and the relation of the aspect ratio and $E/E_F$ is investigated. The described system represents the starting point for the study of novel many-body physics and the universal dynamics with a large-atom number quantum degenerate Fermi gas.

This work is supported by the National Natural Science Foundation of China under Grant No. 11374101 and the Shanghai Pujiang Program under Grant No. 13PJ1402500.


\end{document}